\documentclass[12pt,a4paper]{article}
\usepackage{amsmath,amsfonts,amssymb}
\begin{document}

\title{Exact Solutions in Poincar\'e Gauge Gravity Theory}

\author{Yuri N. Obukhov\\
\small{Institute for Nuclear Safety, Russian Academy of Sciences, 115191 Moscow, Russia}\\ \small{obukhov@ibrae.ac.ru}}

\maketitle

\begin{abstract}In the framework of the gauge theory based on the Poincar\'e symmetry group, the gravitational field is described in terms of the coframe and the local Lorentz connection. Considered as gauge field potentials, they give rise to the corresponding field strength which are naturally identified with the torsion and the curvature on the Riemann--Cartan spacetime. We study the class of quadratic Poincar\'e gauge gravity models with the most general Yang--Mills type Lagrangian which contains all possible parity-even and parity-odd invariants built from the torsion and the curvature. Exact vacuum solutions of the gravitational field equations are constructed as a certain deformation of de Sitter geometry. They are black holes with nontrivial torsion.
\end{abstract}

% Keywords
% \keyword{gauge gravity theory; Poincar\'e group; coframe; Lorentz connection; odd parity}

%%%%%%%%%%%%%%%%%%%%%%%%%%%%%%%%%%%%%%%%%%
\section{Introduction}
%%%%%%%%%%%%%%%%%%%%%%%%%%%%%%%%%%%%%%%%%%

The gauge-theoretic understanding of the fundamental physical interactions is one of the solid cornerstones of the modern science. In simple terms, the gauge principle relates physical forces to the underlying symmetry groups. The corresponding Yang--Mills type formalism was developed for the {\em internal symmetries}, which form the foundation for the theories of electromagnetic, weak and strong interactions, and consistently generalized to the {\em spacetime symmetries} \cite{Reader,PBO} which give rise to the gravitational interaction. In the current research, much attention is paid to the gauge-theoretic models based on the Poincar\'e group, see \cite{Selected,Tartu} for an introduction. The monograph \cite{PBO} provides an extensive list of references. 

It is now well established that Einstein's general relativity (GR) theory provides a valid description of the gravitational phenomena on macroscopic scales. Compared to GR, the gauge gravity theory is expected to improve our understanding of the gravitational physics at microscopic scales (and, likewise, at an early stage of the cosmological evolution of the universe), giving rise to GR in a certain macroscopic limit. The simplest version of the Poincar\'e gauge gravity, known as the Einstein--Cartan theory, is a viable gravitational model which is consistent with all experimental tests, and it only deviates from GR at extremely high matter densities ${\frac {2m^2c^4}{\pi G\hbar^2}}$ (with $m$ mass of a fermion), where it predicts an avoidance of the cosmological singularity \cite{Tartu}. As one knows, the quantized GR is non-renormalizable, and taking into account the success of the Yang--Mills gauge approach for the strong and electro-weak interactions, one hopes that a development of the gauge-theoretic framework for gravity may help in constructing a consistent quantum theory of the gravitational field.

We focus here on the class of Poincar\'e gauge gravity models based on the general quadratic Lagrangians of the Yang--Mills type. Both the parity-even and parity-odd terms are included, extending the previous studies which were mainly confined to the parity symmetric theories. Construction of exact solutions of the field equations is important for checking the validity of the theory, and its consistency with GR and experiments. Here, we report on the black hole solutions with dynamical torsion. 

Our basic notation and conventions are consistent with \cite{MAG,Birk}. In particular, Greek indices $\alpha, \beta, \dots = 0, \dots, 3$, denote the anholonomic components (for example, of a coframe $\vartheta^\alpha$), while the Latin indices $i,j,\dots =0,\dots, 3$, label the holonomic components (e.g., $dx^i$). The anholonomic vector frame basis $e_\alpha$ is dual to the coframe basis in the sense that $e_\alpha\rfloor\vartheta^\beta = \delta_\alpha^\beta$, where $\rfloor$ denotes the interior product. The volume 4-form is denoted by $\eta$, and the $\eta$-basis in the space of exterior forms is constructed with the help of the interior products as $\eta_{\alpha_1 \dots\alpha_p}:= e_{\alpha_p}\rfloor\dots e_{\alpha_1}\rfloor\eta$, $p=1,\dots,4$. They are related to the $\vartheta$-basis via the Hodge dual operator $^*$, for example, $\eta_{\alpha\beta} = {}^*\!\left(\vartheta_\alpha\wedge\vartheta_\beta\right)$. The Minkowski metric is $g_{\alpha\beta} = {\rm diag}(c^2,-1,-1,-1)$. We do not use special unit systems and, accordingly, do not put the fundamental physical constants (such as the light velocity $c$, Planck's constant $\hbar$, and Newton's gravitational constant $G$) equal to one, thereby keeping for all objects their natural dimensions. All the objects related to the parity-odd sector (coupling constants, irreducible pieces of the curvature, etc.) are marked by an overline, to distinguish them from the corresponding parity-even objects.

%%%%%%%%%%%%%%%%%%%%%%%%%%%%%%%%%%%%%%%%%%%%%%%%%
\section{Formal Structure of Poincar\'e Gauge Gravity}\label{PG}
%%%%%%%%%%%%%%%%%%%%%%%%%%%%%%%%%%%%%%%%%%%%%%%%%

The 10-parameter Poincar\'e symmetry group $G = T_4 \rtimes SO(1,3)$ is a semidirect product of the 4-parameter group of translations and the 6-parameter local Lorentz group, and the Yang--Mills--Utiyama--Kibble formalism can be consistently developed on a spacetime manifold \cite{Reader,PBO}. The corresponding gravitational field potentials (``translational'' and ``rotational'', respectively) are then naturally identified with the 1-forms of the coframe and the local Lorentz connection:
\begin{eqnarray}
\vartheta^\alpha &=& e^\alpha_i dx^i,\label{cof}\\
\Gamma^{\alpha\beta} =  -\,\Gamma^{\beta\gamma} &=& \Gamma_{i}{}^{\alpha\beta} dx^i.\label{con}
\end{eqnarray}

The ``translational'' and ``rotational'' field strength 2-forms 
\begin{eqnarray}
T^\alpha &=& D\vartheta^\alpha = d\vartheta^\alpha +\Gamma_\beta{}^\alpha\wedge
\vartheta^\beta,\label{Tor}\\ \label{Cur}
R^{\alpha\beta} &=& d\Gamma^{\alpha\beta} + \Gamma_\gamma{}^\beta\wedge\Gamma^{\alpha\gamma},
\end{eqnarray}
have standard geometrical interpretation as the torsion and the curvature of the Riemann--Cartan spacetime. As usual, the covariant differential is denoted $D$.

%%%%%%%%%%%%%%%%%%%%%%%%%%%%%%%%%%%%%%%%%%
\subsection{Gravitational Field Equations}
%%%%%%%%%%%%%%%%%%%%%%%%%%%%%%%%%%%%%%%%%%

The gravitational Lagrangian 4-form is quite generally an arbitrary invariant function of the geometrical variables:
\begin{equation}
V = V(\vartheta^{\alpha}, T^{\alpha}, R^{\alpha\beta}).\label{lagrV}
\end{equation}

Its variation, with respect to the gravitational (translational and Lorentz) potentials, yields the field~equations
\begin{eqnarray}
{\frac{\delta V}{\delta\vartheta^{\alpha}}} = - DH_{\alpha} + E_{\alpha} &=& 0, \label{dVt}\\ 
{\frac{\delta V}{\delta\Gamma^{\alpha\beta}}}
= - DH_{\alpha\beta} + E_{\alpha\beta} &=& 0.\label{dVG}
\end{eqnarray}

Here, the Poincar\'e {\em gauge field momenta} 2-forms are introduced by
\begin{equation} 
H_{\alpha} := -\,{\frac{\partial V}{\partial T^{\alpha}}}\,,\qquad  
H_{\alpha\beta} := -\,{\frac{\partial V}{\partial R^{\alpha\beta}}}\,,\label{HH}
\end{equation}  
and the $3$-forms of the {\em canonical}{ energy-momentum} and spin for the gravitational gauge fields are constructed as
\begin{eqnarray} 
E_{\alpha} := {\frac{\partial V}{\partial\vartheta^{\alpha}}} &=& e_{\alpha}\rfloor V 
+ (e_{\alpha}\rfloor T^{\beta})\wedge H_{\beta}
+ \,(e_{\alpha}\rfloor R_{\beta}{}^{\gamma})\wedge H^{\beta}{}_{\gamma},\label{Ea}\\
E_{\alpha\beta} := {\frac{\partial V}{\partial\Gamma^{\alpha\beta}}} &=&  
- \,\vartheta_{[\alpha}\wedge H_{\beta]}\,. \label{EE}
\end{eqnarray}

The field Equations (\ref{dVt}) and (\ref{dVG}) are written here for the {\em vacuum case}. In the presence of matter, the right-hand sides of {Equations} (\ref{dVt}) and (\ref{dVG}) contain the canonical energy-momentum and the canonical spin currents of the physical sources, respectively. 

%%%%%%%%%%%%%%%%%%%%%%%%%%%%%%%%%%%%%%%%%%
\subsection{Quadratic Poincar\'e Gravity Models} 
%%%%%%%%%%%%%%%%%%%%%%%%%%%%%%%%%%%%%%%%%%

The torsion 2-form can be decomposed into the 3 irreducible parts, whereas the curvature 2-form has 6 irreducible pieces. Their definition is presented in Appendix~\ref{appA}.

The general quadratic model is described by the Lagrangian 4-form that contains all possible linear and quadratic invariants of the torsion and the curvature:
\begin{eqnarray}
V &=& {\frac {1}{2\kappa c}}\Big\{\Big(a_0\eta_{\alpha\beta} + \overline{a}_0
\vartheta_\alpha\wedge\vartheta_\beta\Big)\wedge R^{\alpha\beta} - 2\lambda_0\eta
-\,T^\alpha\wedge\sum_{I=1}^3\left[a_I\,{}^*({}^{(I)}T_\alpha) + \overline{a}_I
\,{}^{(I)}T_\alpha\right]\Big\} \nonumber\\ &&
- \,{\frac 1{2\rho}}R^{\alpha\beta}\wedge\sum_{I=1}^6 \left[b_I\,{}^*({}^{(I)}\!R_{\alpha\beta}) 
+ \overline{b}_I\,{}^{(I)}\!R_{\alpha\beta}\right].\label{LRT}
\end{eqnarray}

This Lagrangian has a clear structure: The first line encompasses the terms {\em linear} in the curvature and the {\em torsion quadratic} terms (all of which have the same dimension of an area $[\ell^2]$), whereas the second line contains the {\it curvature quadratic} invariants. For completeness, the cosmological constant is included (with the dimension of an inverse area, $[\lambda_0] = [\ell^{-2}]$). Furthermore, each line is composed of parity-even pieces and parity-odd parts (with the coupling constants marked by an overline). A~special case $a_0 = 0$ and $\overline{a}_0 = 0$ describes the purely quadratic model without the Hilbert--Einstein linear term in the Lagrangian. 

A general Poincar\'e gauge model contains a set of the coupling constants which determine the structure of the quadratic part of the Lagrangian: $\rho$, $a_1, a_2, a_3$, $b_1, \cdots, b_6$ and $\overline{a}_1, \overline{a}_2, \overline{a}_3$, $\overline{b}_1, \cdots, \overline{b}_6$. It is important to note that not all of the latter constants are independent---we take $\overline{a}_2 = \overline{a}_3$, $\overline{b}_2 = \overline{b}_4$ and $\overline{b}_3 = \overline{b}_6$ because some of the terms in {Lagrangian} (\ref{LRT}) are the same in view of {Equations} (\ref{T11})--(\ref{R36}). As we already mentioned, the overbar denotes the constants responsible for the parity-odd sector. In recent times, there is a growing interest to such interactions \cite{OPZ,HoN,Diakonov,Baekler1,Baekler2,Kara,BC}. Quite generally, there are no compelling theoretical arguments or experimental evidence which could rule out the violation of parity in gravity, and in 1964 Leitner and Okubo \cite{LO} looked into a possibility of extending the gravitational Lagrangian by parity odd terms. Later such extensions were widely studied in the context of the classical and quantum gravity theory \cite{Hari,Hoj}, in particular in Ashtekar's approach and loop quantum gravity \cite{Holst,CK}. Moreover, the inclusion of parity-nonconserving terms is important for the discussion of such fundamental physical issues as the baryon asymmetry of the universe, where the parity-odd terms can be induced by the quantum vacuum structure \cite{Randono,Pop,Bj}.

All coupling constants $a_I$, $\overline{a}_I$, $b_I$, and $\overline{b}_I$ are dimensionless, whereas the dimension $[{\frac 1\rho}] = [\hbar]$. Keeping in mind the importance of the macroscopic limit to GR, we have $\kappa = 8\pi G/c^4$ as Einstein's gravitational constant. The microscopic gravitational phenomena are naturally characterized by the~parameter 
\begin{equation}
\ell_\rho^2 = {\frac {\kappa c}{\rho}}.\label{lr}
\end{equation}

Since $[{\frac 1\rho}] = [\hbar]$, this new coupling parameter has the dimension of an area, $[\ell_\rho^2] = [\ell^2]$. Below we will see that $\ell_\rho^2$ parameter describes the contribution of the curvature square terms in the Lagrangian (\ref{LRT}) to the gravitational field dynamics in Equations (\ref{ERT1}) and (\ref{ERT2}). 

For the Lagrangian (\ref{LRT}), from  {Equations} (\ref{HH})--(\ref{EE}) we derive the gauge gravitational field momenta 
\begin{eqnarray}
H_\alpha = {\frac 1{\kappa c}}\,h_\alpha\,,\qquad
H_{\alpha\beta} = -\,{\frac {1}{2\kappa c}}\left(a_0\,\eta_{\alpha\beta} + \overline{a}_0
\vartheta_\alpha\wedge\vartheta_\beta\right)  + {\frac 1\rho}\,h_{\alpha\beta},\label{HabRT}
\end{eqnarray}
and the canonical energy-momentum and spin currents of the gravitational field
\begin{eqnarray}
E_\alpha = {\frac {1}{2\kappa c}}\Big(a_0\,\eta_{\alpha\beta\gamma}\wedge R^{\beta\gamma} + 
2\overline{a}_0\,R_{\alpha\beta}\wedge\vartheta^\beta - 2\lambda_0\eta_\alpha +
q^{(T)}_\alpha\Big) + {\frac 1\rho}\,q^{(R)}_\alpha,\qquad 
E_{\alpha\beta} = {\frac 1{\kappa c}}\,h_{[\alpha}\wedge\vartheta_{\beta]}.\label{EabRT}
\end{eqnarray}

For convenience, we introduced here the 2-forms that are linear functions 
of the torsion and the curvature, respectively, by
\begin{eqnarray}
h_\alpha = \sum_{I=1}^3\left[a_I\,{}^*({}^{(I)}T_\alpha) 
+ \overline{a}_I\,{}^{(I)}T_\alpha\right],\qquad 
h_{\alpha\beta} = \sum_{I=1}^6\left[b_I\,{}^*({}^{(I)}\!R_{\alpha\beta}) 
+ \overline{b}_I\,{}^{(I)}\!R_{\alpha\beta}\right],\label{hR}
\end{eqnarray}
and the 3-forms which are quadratic in the torsion and in the curvature, respectively:
\begin{eqnarray}
q^{(T)}_\alpha = {\frac 12}\left[(e_\alpha\rfloor T^\beta)\wedge h_\beta - T^\beta\wedge 
e_\alpha\rfloor h_\beta\right],\qquad 
q^{(R)}_\alpha = {\frac 12}\left[(e_\alpha\rfloor R^{\beta\gamma})\wedge h_{\beta\gamma} 
- R^{\beta\gamma}\wedge e_\alpha\rfloor h_{\beta\gamma}\right].\label{qaR}
\end{eqnarray}

By construction, $[h_\alpha] = [\ell]$ has the dimension of a length, whereas the 2-form $[h_{\alpha\beta}] = 1$ is obviously dimensionless. Similarly, we find for {Equations} (\ref{qaR}) the dimension of length $[q^{(T)}_\alpha] = [\ell]$, and the dimension of the inverse length, $[q^{(R)}_\alpha] = [1/\ell]$, respectively. 

The resulting {\it vacuum} Poincar\'e gravity field Equations (\ref{dVt}) and (\ref{dVG}) then read:
\begin{eqnarray}
{\frac {a_0}2}\eta_{\alpha\beta\gamma}\wedge R^{\beta\gamma} + \overline{a}_0R_{\alpha\beta}
\wedge\vartheta^\beta - \lambda_0\eta_\alpha + q^{(T)}_\alpha + \ell_\rho^2\,q^{(R)}_\alpha
- Dh_\alpha &=& 0,\label{ERT1}\\
{\frac {a_0}2}\,\eta_{\alpha\beta\gamma}\wedge T^{\gamma} + \overline{a}_0
\,T_{[\alpha}\wedge\vartheta_{\beta]} + h_{[\alpha}\wedge\vartheta_{\beta]} 
- \ell_\rho^2\,Dh_{\alpha\beta} &=& 0.\label{ERT2}
\end{eqnarray}

%%%%%%%%%%%%%%%%%%%%%%%%%%%%%%%%%%%%%%%%%%
\section{Prelude: de Sitter Spacetime}
%%%%%%%%%%%%%%%%%%%%%%%%%%%%%%%%%%%%%%%%%%

As a preliminary step, we discuss the de Sitter spacetime in an unusual disguise.
This manifold has many faces, and here we consider one of them which is not quite well known. Using a spherical local coordinate system $(t,r,\theta,\varphi)$, it is given by the coframe
\begin{align}\label{cofDS}
\widehat{\vartheta}^\alpha = 
\begin{cases}
\widehat{\vartheta}{}^{\hat 0} =& {\frac {\Delta + mr}{\Delta}}\,\sqrt{\frac{\Delta}{\Sigma}}\left[cdt - j_0\sin^2\theta\,d\varphi\right] + {\frac {mr}{\Delta}}\,\sqrt{\frac{\Sigma}{\Delta}}\,dr,\\ 
\widehat{\vartheta}{}^{\hat 1} =& {\frac {mr}{\Delta}}\,\sqrt{\frac{\Delta}{\Sigma}}\left[cdt - j_0\sin^2\theta\,d\varphi\right] + {\frac {\Delta - mr}{\Delta}}\,\sqrt{\frac{\Sigma}{\Delta}}\, dr,\\ 
\widehat{\vartheta}{}^{\hat 2} =& \sqrt{\frac{\Sigma}{f}}\,d\theta,\\
\widehat{\vartheta}{}^{\hat 3} =& \sqrt{\frac{f}{\Sigma}}\sin\theta\left[ - j_0\,cdt + %\Omega
(r^2 + j_0^2)\,d\varphi\right].
\end{cases}
\end{align}

Here, the rotation parameter is denoted by $j_0$ (in order to distinguish it from the torsion coupling constants we avoid a more common notation $a$), and the functions and constants are defined by
\begin{eqnarray}
\Delta&:=& (r^2 + j_0^2)(1 - \lambda\,r^2) - 2mr,\\
\Sigma&:=& r^2 + j_0^2\cos^2\theta,\\
f &:=& 1 + \lambda\,j_0^2\cos^2\theta,\\
m &:=& \frac{GM}{c^2},
\end{eqnarray}
and $0<t<\infty$, $0<r<\infty$, $0<\theta<\pi$ and $0<\varphi<2\pi$.

The corresponding line element $ds^2 = g_{\alpha\beta}\widehat{\vartheta}{}^\alpha\otimes\widehat{\vartheta}{}^\beta = \widehat{g}_{ij}dx^idx^j$, with the spacetime metric $\widehat{g}_{ij} = \widehat{e}{}_i^\alpha \widehat{e}{}_j^\beta g_{\alpha\beta}$, reads:
\begin{eqnarray}
ds^2 &=& \left[f - \lambda (r^2 + j_0^2)\right]c^2dt^2 - {\frac {4mr}{\Delta}}\,cdt\,dr \nonumber + 2\lambda\,(r^2 + j_0^2)\,j_0\sin^2\!\theta\,cdt\,d\varphi \\
&& +\,{\frac {4mr}{\Delta}}\,j_0\sin^2\!\theta\,dr\,d\varphi - {\frac {\Delta - 2mr}{\Delta^2}}\,\Sigma\,dr^2 - {\frac {\Sigma}{f}}\,d\theta^2 - (1 + \lambda j_0^2)(r^2 + j_0^2)\,\sin^2\!\theta\,d\varphi^2.\label{dsADS}
\end{eqnarray}

When $m = 0$ and $j_0 = 0$, this line element reduces to
\begin{eqnarray}
ds^2 = (1 - \lambda r^2)\,c^2dt^2 - {\frac {dr^2}{1 - \lambda r^2}}
- r^2(d\theta^2 + \sin^2\!\theta\,d\varphi^2),\label{dsK}
\end{eqnarray}
which represents the static spherically symmetric form of the de Sitter spacetime. Quite remarkably, however, also in the general case with $m \neq 0$ and $j_0 \neq 0$, despite a complicated form of the coframe (\ref{cofDS}) and the metric (\ref{dsADS}), the components of which appear to be highly nontrivial functions of the spacetime coordinates and parameters $m, j_0, \lambda$, the corresponding Riemannian connection satisfies
\begin{eqnarray}
\widehat{T}{}^\alpha &=& d\widehat{\vartheta}{}^\alpha + \widehat{\Gamma}_\beta{}^\alpha\wedge\widehat{\vartheta}{}^\beta = 0,\label{DSnoT}\\
\widehat{R}{}^{\alpha\beta} &=& d\widehat{\Gamma}{}^{\alpha\beta} + \widehat{\Gamma}{}_\gamma{}^\beta\wedge\widehat{\Gamma}{}^{\alpha\gamma} = \lambda\,\widehat{\vartheta}{}^\alpha\wedge\widehat{\vartheta}{}^\beta.\label{DScur}
\end{eqnarray}

The components of the connection $\widehat{\Gamma}{}^{\alpha\beta}$ are explicitly given in Appendix~\ref{appB}. By making use of {Equations} (\ref{Gds1})--(\ref{Gds3}), it is straightforward (although the corresponding computation is somewhat lengthy) to verify {Equation} (\ref{DScur}). 

In other words, even for nonvanishing $m$ and $j_0$, the coframe (\ref{cofDS}) and the metric (\ref{dsADS}) describe the torsionless de Sitter spacetime of the constant curvature $\lambda$. Note that depending on the sign of $\lambda$, one sometimes speaks of de Sitter and anti-de Sitter geometries. Here, we do not use this refined language and---irrespective of the sign of the constant curvature---call all these spaces de Sitter.

%%%%%%%%%%%%%%%%%%%%%%%%%%%%%%%%%%%%%%%%%%
\section{Interlude: From de Sitter to Kerr--de Sitter}
%%%%%%%%%%%%%%%%%%%%%%%%%%%%%%%%%%%%%%%%%%

As a next step, we introduce the 1-form
\begin{equation}
k := \sqrt{\frac {\Sigma}{\Delta}}\left(\widehat{\vartheta}{\,}^{\hat{0}} -
\widehat{\vartheta}{\,}^{\hat{1}}\right) = cdt - {\frac {\Sigma}{\Delta}}\,dr 
- j_0\,\sin^2\theta\,d\varphi,\label{k_ds}
\end{equation}
and define the covector components by $k_\alpha = \widehat{e}_\alpha\rfloor k$. One can straightforwardly check that this construction yields a null geodetic congruence:
\begin{equation}\label{k_null}
k\wedge{}^{\widehat{\ast}} k = 0,\qquad k\wedge{}^{\widehat{\ast}}\widehat{D}k_\alpha = 0.
\end{equation}

Hereafter the hat marks the objects and operators in the de Sitter space ($\widehat{\vartheta}^\alpha$, $\widehat{\Gamma}{}^{\alpha\beta}$). Namely, $\widehat{D}$ is the covariant differential corresponding to the connection $\widehat{\Gamma}{}^{\alpha\beta}$, and ${}^{\widehat{\ast}}$ is the Hodge duality operator corresponding to the coframe $\widehat{\vartheta}^\alpha$. 

Now we use the null 1-form (\ref{k_ds}) to define a new coframe
\begin{equation}
\vartheta^\alpha = \widehat{\vartheta}{}^\alpha - Uk^\alpha k,\label{cofkerr}
\end{equation}
which can be considered as a kind of perturbation of the de Sitter coframe (\ref{cofDS}). The components of the modified coframe $\vartheta^\alpha = e^\alpha_idx^i$ are now $e^\alpha_i = \widehat{e}{}^\alpha_i - Uk^\alpha k_i$, and the corresponding spacetime metric has the Kerr--Schild form
\begin{equation}
g_{ij} = e^\alpha_i e^\beta_jg_{\alpha\beta} = \widehat{g}{}_{ij} - 2Uk_ik_j.\label{metkerr}
\end{equation}
  
Here $U = U(r,\theta)$ is a function of the radial coordinate $r$ and the angle $\theta$. In order to preserve the stationarity and the axial symmetry, we assume that all the geometric variables do not depend on $t$ and $\varphi$. If we choose
\begin{equation}
U = {\frac {mr}{\Sigma}},\label{Ukerr}
\end{equation}
the coframe (\ref{cofkerr}) reads explicitly 
\begin{align}\label{cofK}
\vartheta^\alpha =
\begin{cases}  
\vartheta^{\hat 0} =& \sqrt{\frac{\Delta}{\Sigma}}\left[cdt - j_0 %\Omega
\sin^2\theta\,d\varphi\right],\\
\vartheta^{\hat 1} =& \sqrt{\frac{\Sigma}{\Delta}}\, dr, \\
\vartheta^{\hat 2} =& \sqrt{\frac{\Sigma}{f}}\, d\theta, \\
\vartheta^{\hat 3} =& \sqrt{\frac{f}{\Sigma}}\sin\theta\left[ - j_0\,cdt + %\Omega
(r^2 + j_0^2)\,d\varphi\right].
\end{cases}
\end{align}

We immediately recognize the coframe of the {\em Kerr--de Sitter geometry} \cite{Kerr}.

%%%%%%%%%%%%%%%%%%%%%%%%%%%%%%%%%%%%%%%%%%
\section{Postlude: Ansatz for Poincar\'e Gauge Theory}
%%%%%%%%%%%%%%%%%%%%%%%%%%%%%%%%%%%%%%%%%%

In Poincar\'e gauge theory, the ansatz for the translational potential (coframe 1-form) should be supplemented by the ansatz for the local Lorentz connection 1-form. After all the preparations, we are now in a position to formulate the {\it Baekler ansatz} for the Poincar\'e gauge fields:
\begin{eqnarray}
\vartheta^\alpha &=& \widehat{\vartheta}{}^\alpha - Uk^\alpha k,\label{cofB}\\
\Gamma^{\alpha\beta} &=& \widehat{\Gamma}{}^{\alpha\beta}.\label{gamB}  
\end{eqnarray}

Earlier, a similar technique was successfully used for the construction of the exact plane wave solutions in the Poincar\'e gauge theory \cite{PGW1,PGW2}.

%%%%%%%%%%%%%%%%%%%%%%%%%%%%%%%%%%%%%%%%%%
\section{Solving Gravitational Field Equations}
%%%%%%%%%%%%%%%%%%%%%%%%%%%%%%%%%%%%%%%%%%

It is important to notice that the null 1-form (\ref{k_ds}) preserves its structure with respect to the new coframe (\ref{cofkerr}):
\begin{equation}
k = \sqrt{\frac {\Sigma}{\Delta}}\left(\vartheta^{\hat{0}} -
\vartheta^{\hat{1}}\right),\label{kB}
\end{equation}
and consequently the covector components $k_\alpha = e_\alpha\rfloor k = \sqrt{\frac {\Sigma}{\Delta}}(1, -1, 0, 0)$ have the same values. Moreover, it is still a null geodetic congruence,
\begin{equation}\label{knullB}
k\wedge{}^{\ast} k = 0,\qquad k\wedge{}^{\ast}Dk_\alpha = 0.
\end{equation}

It is straightforward to derive the torsion and the Riemann--Cartan curvature for the Baekler ansatz (\ref{cofB}) and (\ref{gamB}). Combining {Equations} (\ref{gamB}), (\ref{DScur}), and (\ref{cofB}) we find
\begin{eqnarray}\label{torB}
T^\alpha &=& D\vartheta^\alpha = \widehat{D}\vartheta^\alpha = -\,\widehat{D}(Uk^\alpha k),\\
R^{\alpha\beta} &=& \widehat{R}{}^{\alpha\beta} = \lambda\,\widehat{\vartheta}{}^\alpha\wedge
\widehat{\vartheta}{}^\beta = \lambda\,\vartheta^\alpha\wedge\vartheta^\beta + 2\lambda U k\wedge k^{[\alpha}\vartheta^{\beta]}.\label{curB}
\end{eqnarray}

One can check the following properties of the Poincar\'e gauge field strengths:
\begin{eqnarray}\label{kTR}
k_\alpha T^\alpha = 0,&\qquad& k_\alpha (R^{\alpha\beta} - \lambda\vartheta^\alpha\wedge\vartheta^\beta) = 0,\\
\vartheta_\alpha\wedge T^\alpha = 0,&\qquad& \vartheta_\alpha\wedge R^{\alpha\beta} = 0.\label{vTR}
\end{eqnarray}

Next, we need to find the irreducible parts. Using the identities (\ref{kTR}) and (\ref{vTR}), we can verify that most of the irreducible parts of the curvature vanish, except for the 4th and 6th:
\begin{eqnarray}\label{R12}
{}^{(1)}R^{\alpha\beta} &=& {}^{(2)}R^{\alpha\beta} = {}^{(3)}R^{\alpha\beta} = {}^{(5)}R^{\alpha\beta} = 0,\\
{}^{(4)}R^{\alpha\beta} &=& 2\lambda U k\wedge k^{[\alpha}\vartheta^{\beta]},\qquad
{}^{(6)}R^{\alpha\beta} = \lambda\,\vartheta^\alpha\wedge\vartheta^\beta.\label{R46}
\end{eqnarray}

As compared to the curvature (\ref{curB}), the structure of the torsion (\ref{torB}) is more nontrivial. Making use of the components of the connection ({Equations} (\ref{Gds1})--(\ref{Gds3})) and {Equation} (\ref{Ukerr}), one can evaluate the covariant derivative in~(\ref{torB}) to get 
\begin{equation}\label{torB0}
T^\alpha = -\,v_5\,k\wedge\vartheta^\alpha + v_4{}^\ast(k\wedge\vartheta^\alpha) + k^\alpha w,
\end{equation}
where we introduce the 2-form
\begin{equation}
w = (v_1 + v_5)\,\vartheta^{\hat 0}\wedge\vartheta^{\hat 1} + 3v_4\,\vartheta^{\hat 2}\wedge
\vartheta^{\hat 3} - k\wedge(v_2\vartheta^{\hat 2} + v_3\vartheta^{\hat 3}),\label{wkerr}
\end{equation}
and denoted the functions
\begin{eqnarray}
v_1 &=& -\,{\frac {m(r^2 - j_0^2\cos^2\theta)}{\Sigma^2}},\quad v_2 =
\sqrt{\frac f\Sigma}\,{\frac {mrj_0^2\sin\theta\cos\theta}{\Sigma^2}},\label{v145}\\
v_3 &=& \sqrt{\frac f\Sigma}\,{\frac {mr^2j_0\sin\theta}{\Sigma^2}},\quad
v_4 = {\frac {mrj_0\cos\theta}{\Sigma^2}},\quad v_5 = -\,{\frac {mr^2}{\Sigma^2}}.\label{v23}
\end{eqnarray}

Explicitly, the components of the torsion 2-form (\ref{torB0}) read:
\begin{align}\label{TBa}
T^\alpha = 
\begin{cases}
T^{\hat 0} =& k^{\hat 0}\,[ v_1\vartheta^{\hat 0}
\wedge\vartheta^{\hat 1} + 2v_4\,\vartheta^{\hat 2}\wedge\vartheta^{\hat 3}
- k\wedge(v_2\vartheta^{\hat 2} + v_3\vartheta^{\hat 3})],\\
T^{\hat 1} =& k^{\hat 1}\,[ v_1\vartheta^{\hat 0}
\wedge\vartheta^{\hat 1} + 2v_4\,\vartheta^{\hat 2}\wedge\vartheta^{\hat 3}
- k\wedge(v_2\vartheta^{\hat 2} + v_3\vartheta^{\hat 3})],\\
T^{\hat 2} =& -\,k\wedge (v_5\vartheta^{\hat 2} + v_4\vartheta^{\hat 3}),\\
T^{\hat 3} =& -\,k\wedge ( -\,v_4\vartheta^{\hat 2} + v_5\vartheta^{\hat 3}).
\end{cases}
\end{align}

One can verify that the torsion trace is proportional to the 1-form $k$:
\begin{equation}
T = e_\alpha\rfloor T^\alpha = (2v_5 - v_1)\,k = -\,{\frac {m}{\Sigma}}\,k.\label{Ttrace}
\end{equation}

As a result, we find the irreducible parts of the torsion: ${}^{(3)}T^\alpha = 0$ and
\begin{eqnarray}
{}^{(1)}T^\alpha &=& -\,{\frac {(v_1 + v_5)}3}\,k\wedge\vartheta^\alpha
+ v_4{}^\ast(k\wedge\vartheta^\alpha) + k^\alpha w,\label{tor1B}\\
{}^{(2)}T^\alpha &=& {\frac {(v_1 - 2v_5)}3}\,k\wedge\vartheta^\alpha.\label{tor2B}
\end{eqnarray}

The Riemann--Cartan geometry of Baekler's configuration (\ref{cofB}) and (\ref{gamB}) is globally {\it regular} in the sense that all the torsion invariants vanish, whereas curvature invariants are constant. In particular,
\begin{equation}\label{invB}
T^\alpha\wedge{}^\ast\!T_\alpha = 0,\qquad R^{\alpha\beta}\wedge{}^\ast\!R_{\alpha\beta} = 12\lambda^2\eta.
\end{equation}

To solve the coupled system of the Poincar\'e gauge field Equations (\ref{ERT1}) and (\ref{ERT2}), we have to evaluate $q^{(T)}_\alpha$, $q^{(R)}_\alpha$, $h_\alpha$, $h_{\alpha\beta}$, and the covariant derivative $Dh_\alpha$, $Dh_{\alpha\beta}$. We begin by noticing that the structure of the 2-form $h_{\alpha\beta}$ realizes the generalized double-duality ansatz \cite{Selected}, namely:
\begin{equation}
h_{\alpha\beta} = \lambda_1\,{\frac 12}\eta_{\alpha\beta\mu\nu}R^{\mu\nu} + 
\lambda_2\,R_{\alpha\beta} + \lambda_3\,\eta_{\alpha\beta} + \lambda_4
\,\vartheta_\alpha\wedge\vartheta_\beta.\label{DDh}
\end{equation}

In view of {Equations} (\ref{R12}) and (\ref{R46}), we have explicitly
\begin{equation}\label{llbb}
\lambda_1 = -\,b_4,\quad \lambda_2 = \overline{b}{}_4,\quad \lambda_3 =
(b_4 + b_6)\lambda,\quad \lambda_4 = -\,(\overline{b}{}_4 - \overline{b}{}_6)\lambda.
\end{equation}

Making use of {Equation} (\ref{DDh}), we find
\begin{equation}
q^{(R)}_\alpha = -\,\lambda_3\eta_{\alpha\beta\gamma}\wedge R^{\beta\gamma} - 2\lambda_4
R_{\alpha\beta}\wedge\vartheta^\beta + 6\lambda\lambda_3\eta_\alpha,\label{qRDDA}
\end{equation}
and recast the field equations (\ref{ERT1}) and (\ref{ERT2}) into
\begin{eqnarray}
{\frac {a_{\rm eff}}2}\eta_{\alpha\beta\gamma}\wedge R^{\beta\gamma} + \overline{a}{}_{\rm eff}
R_{\alpha\beta}\wedge\vartheta^\beta - \lambda_{\rm eff}\eta_\alpha + q^{(T)}_\alpha - Dh_\alpha  
&=& 0,\label{EDD1}\\
{\frac {a_{\rm eff}}2}\,\eta_{\alpha\beta\gamma}\wedge T^{\gamma} + \overline{a}{}_{\rm eff}
\,T_{[\alpha}\wedge\vartheta_{\beta]} + h_{[\alpha}\wedge\vartheta_{\beta]} &=& 0,\label{EDD2}
\end{eqnarray}
where we introduce the effective constants
\begin{eqnarray}
a_{\rm eff} = a_0 - 2\ell_\rho^2\lambda_3,\qquad
\overline{a}{}_{\rm eff} = \overline{a}{}_0 - 2\ell_\rho^2\lambda_4,\qquad
\lambda_{\rm eff} = \lambda_0 - 6\ell_\rho^2\lambda_3\lambda.\label{lEff}
\end{eqnarray}

In order to simplify the first field equation (\ref{EDD1}), we use the explicit form of the curvature (\ref{curB}) to find for the first term
\begin{eqnarray}\label{eR}
{\frac {1}2}\eta_{\alpha\beta\gamma}\wedge R^{\beta\gamma} = {\frac {\lambda}2}
\,\eta_{\alpha\beta\gamma}\wedge \vartheta^\beta\wedge\vartheta^\gamma +
\lambda U\eta_{\alpha\beta\gamma}\wedge kk^\beta\wedge\vartheta^\gamma 
= 3\lambda\,\eta_\alpha + 2\lambda U k_\alpha k\wedge\vartheta^{\hat 2}\wedge\vartheta^{\hat 3},
\end{eqnarray}
whereas the second term vanishes $R_{\alpha\beta}\wedge\vartheta^\beta = 0$ in view of (\ref{vTR}).

A direct computation yields
\begin{equation}\label{qaB}
q^{(T)}_\alpha = {\frac {2m}{3\Sigma}}\left[(2a_1 + a_2)v_5 - (\overline{a}{}_1
- \overline{a}{}_2)v_4\right]k_\alpha k\wedge\vartheta^{\widehat 2}\wedge\vartheta^{\widehat 3},
\end{equation}
whereas after a long algebra we find for the components of the derivative $Dh_\alpha$:
\begin{eqnarray}
-Dh_{\hat 0} &=& -\,{\frac {2(\overline{a}{}_1 - \overline{a}{}_2)v_4}{3r}}\,
\widehat{\vartheta}{}^{\hat 0}\wedge\vartheta^{\hat 2}\wedge\vartheta^{\hat 3} +   
{\frac {2a_1 + a_2}{3r}}\,k\wedge\vartheta^{\hat 0}\wedge\left(v_3\vartheta^{\hat 2} 
+ v_2\vartheta^{\hat 3}\right)\nonumber\\
&& +\,{\frac {v_5}{3r}}\sqrt{\frac {\Delta}{\Sigma}}\,k\wedge\vartheta^{\hat 2}\wedge
\vartheta^{\hat 3}\left\{-\,6a_1\lambda{\frac {\Sigma^2}{\Delta}} 
+ \,(2a_1 + a_2)\left[ 1 - {\frac {\Sigma}{\Delta}}(1 - \lambda j_0^2
- 2\lambda r^2)\right]\right\},\label{Dh0}\\
-Dh_{\hat 1} &=& {\frac {2(\overline{a}{}_1 - \overline{a}{}_2)v_4}{3r}}\,
\widehat{\vartheta}{}^{\hat 1}\wedge\vartheta^{\hat 2}\wedge\vartheta^{\hat 3} -
{\frac {2a_1 + a_2}{3r}}\,k\wedge\vartheta^{\hat 1}\wedge\left(v_3\vartheta^{\hat 2} 
+ v_2\vartheta^{\hat 3}\right)\nonumber\\
&& +\,{\frac {v_5}{3r}}\sqrt{\frac {\Delta}{\Sigma}}\,k\wedge\vartheta^{\hat 2}\wedge
\vartheta^{\hat 3}\left\{6a_1\lambda{\frac {\Sigma^2}{\Delta}} 
+ \,(2a_1 + a_2)\left[ 1 + {\frac {\Sigma}{\Delta}}(1 - \lambda j_0^2
- 2\lambda r^2)\right]\right\},\label{Dh1}\\
-Dh_{\hat 2} &=& {\frac 1{3r}}\left[(2a_1 + a_2)v_4 + 2(\overline{a}{}_1 - \overline{a}{}_2)
v_5\right]\vartheta^{\hat 0}\wedge\vartheta^{\hat 1}\wedge\vartheta^{\hat 2}\nonumber\\
&& +\,{\frac {2a_1 + a_2}{3r}}\left[v_5\vartheta^{\hat 0}\wedge\vartheta^{\hat 1}\wedge
\vartheta^{\hat 3} + v_2\,k\wedge\vartheta^{\hat 2}\wedge\vartheta^{\hat 3}\right],\label{Dh2}\\
-Dh_{\hat 3} &=& {\frac 1{3r}}\left[(2a_1 + a_2)v_4 + 2(\overline{a}{}_1 - \overline{a}{}_2)
v_5\right]\vartheta^{\hat 0}\wedge\vartheta^{\hat 1}\wedge\vartheta^{\hat 3}\nonumber\\
&& +\,{\frac {2a_1 + a_2}{3r}}\left[-\,v_5\vartheta^{\hat 0}\wedge\vartheta^{\hat 1}\wedge
\vartheta^{\hat 2} + v_3\,k\wedge\vartheta^{\hat 2}\wedge\vartheta^{\hat 3}\right].\label{Dh3}
\end{eqnarray}

As we can see, {Equations} (\ref{qaB})--(\ref{Dh3}) are significantly simplified when the coupling constants satisfy $2a_1 + a_2 = 0$ and $\overline{a}{}_1 = \overline{a}{}_2$. Namely, we then get $q^{(T)}_\alpha = 0$ and, by making use of {Equations} (\ref{v145}) and (\ref{Ukerr}), we find 
\begin{equation}\label{DhB}
-Dh_\alpha = -\,{\frac {6a_1\lambda\,v_5\Sigma}{3r}}\,k_\alpha k\wedge\vartheta^{\hat 2}\wedge
\vartheta^{\hat 3} = 2a_1\lambda\,U\,k_\alpha k\wedge\vartheta^{\hat 2}\wedge\vartheta^{\hat 3}.
\end{equation}

Now let us turn to the second field Equation (\ref{EDD2}) which, in view of the definition (\ref{hR}), imposes an algebraic constraint on the spacetime torsion. Indeed, by making use of the properties of irreducible parts of the torsion, we recast {Equation} (\ref{EDD2}) into an equivalent form
\begin{eqnarray}
-\,(a_{\rm eff} + a_1){}^*({}^{(1)}T^\alpha) + (2a_{\rm eff} - a_2){}^*({}^{(2)}T^\alpha) 
- \left(\overline{a}{}_{\rm eff} +\overline{a}_1\right){}^{(1)}T^\alpha - \left(
\overline{a}{}_{\rm eff} + \overline{a}_2\right){}^{(2)}T^\alpha = 0.\label{EDDT2}
\end{eqnarray}

This admits a nontrivial field configuration when $a_{\rm eff} + a_1 = 0$ and $\overline{a}{}_{\rm eff} + \overline{a}{}_1 = 0$. Combining the definitions (\ref{lEff}) with {Equation} (\ref{llbb}), we then finally obtain the constraints on the coupling constants 
\begin{eqnarray}\label{atb01}
a_0 + a_1 - 2\ell_\rho^2\lambda\,(b_4 + b_6) = 0,\qquad \overline{a}{}_0
+ \overline{a}{}_1 + 2\ell_\rho^2\lambda\,(\overline{b}{}_4 - \overline{b}{}_6)= 0.
\end{eqnarray}

Substituting {Equation} (\ref{DhB}) and {Equation} (\ref{eR}) into {Equation} (\ref{EDD1}), we discover that the 1st field equation reduces to a simple relation $3a_{\rm eff}\lambda = \lambda_{\rm eff}$. Recalling the definitions of the effective coupling constants (\ref{lEff}), the latter is equivalent to 
\begin{equation}
3a_0\lambda = \lambda_0 .\label{alB}
\end{equation}

%%%%%%%%%%%%%%%%%%%%%%%%%%%%%%%%%%%%%%%%%%
\section{Discussion and Conclusions}
%%%%%%%%%%%%%%%%%%%%%%%%%%%%%%%%%%%%%%%%%%

The algebraic constraints on the torsion is a well known feature of the double-duality technique \cite{Selected} which leads to restrictions on the coupling constants. Nevertheless, the class of quadratic theories still remains very wide, and it includes many physically viable models. Moreover, one of the outstanding problems in the Poincar\'e gauge gravity is the search for the physically meaningful conditions that improve the structure of general Lagrangians so as to pass the ``consistency check'' with GR. The~latter means a possibility of a smooth recovery of GR results in a certain macroscopic limit. In particular, the existence of the black hole solutions can be considered a manifestation of such a consistency with GR for the class of models which allow for the exact solutions obtained above.

In this sense, it is worthwhile to mention one of the most interesting Poincar\'e gravity models, which was proposed by von der Heyde \cite{vdh1} and attracted much attention in the early stages of development of the gauge approach in gravitational theory \cite{vdh2}. A peculiar feature of the {\em von der Heyde} Lagrangian is that it does not contain the Hilbert term linear in the curvature, therefore it is a~purely quadratic model both in the torsion and the curvature. Explicitly, this Lagrangian reads 
\begin{eqnarray}\label{vdH}
V^{\rm vdH} = {\frac 1{2\kappa c}}\left\{ (\vartheta^\alpha\wedge T^\beta )\wedge
{}^*(\vartheta_\beta\wedge T_\alpha ) + {\frac {1}{4\lambda}}R^{\alpha\beta}\wedge
{}^*R_{\alpha\beta}\right\}.
\end{eqnarray}

The parity-odd terms are absent, whereas the parity-even sector is described by the set of the coupling constants as $a_0 = 0$, $\lambda_0 = 0$, $b_1 = b_2 = b_3 = b_4 = b_6 = b$, and 
\begin{equation}
a_1 = -\,1,\qquad a_2 = 2,\qquad a_3 = 0,\qquad {\frac {b}{\rho}} = -\,{\frac {1}{4\kappa c\lambda}}.\label{avdH}
\end{equation}

Another feasible gauge gravity model arises in a de Sitter gauge approach when the Poincar\'e symmetry is extended to a 10-parameter de Sitter group \cite{SW,Ark}. The corresponding model is described by the Lagrangian
\begin{eqnarray}
V^{\rm dS} = {\frac 1{2\kappa c}}\left\{R_{\alpha\beta}\wedge\eta^{\alpha\beta} - 6\lambda\eta
- {\frac {1}{4\lambda}}R^{\alpha\beta}\wedge{}^*R_{\alpha\beta}\right\}.\label{VdS}
\end{eqnarray}

The coupling constants set then reduces to $a_0 = 1$, $\lambda_0 = 3\lambda$, $b_1 = b_2 = b_3 = b_4 = b_6 = b$, and 
\begin{equation}
a_1 = a_2 = a_3 = 0,\qquad {\frac {b}{\rho}} = {\frac {1}{4\kappa c\lambda}}.\label{adS}
\end{equation}

As compared to the von der Heyde model (\ref{vdH}), the Lagrangian (\ref{VdS}) does include the explicit Hilbert~term. 

The two gravity theories (\ref{vdH}) and (\ref{VdS}) are explicit examples of the models which satisfy the consistency principle with GR, they both admit the black hole solutions described above. However, the results obtained here are important in view of their widest possible applicability. The quadratic Yang--Mills type Lagrangian (\ref{LRT}) describes the class of the most general Poincar\'e gauge gravity models with the dynamical torsion. The existence of the black hole solutions (with Schwarzschild, Kerr, or Kerr--de Sitter metric) was earlier reported for the case of the parity symmetric theories \cite{PB1,PB2,PB3,PB4,PB5,PB6,PB7,PB8}. Now we demonstrated this for the general case with both parity-even and parity-odd sectors taken into~account.

In addition, we have clarified the underlying geometrical structure of these exact solutions. Namely, in the framework of the first-order formalism with the Poincar\'e gauge potentials $(\vartheta^\alpha, \Gamma^{\alpha\beta})$ as the fundamental field variables (as compared to the second-order formalism used in \cite{PB1,PB2,PB3,PB4,PB5,PB6,PB7,PB8} with the metric and torsion as the basic variables), the black hole solution is constructed with the help of the beautiful ansatz (\ref{cofB}) and (\ref{gamB}). Thereby, we have essentially developed a generalization to the post-Riemannian geometries of the Kerr--Schild technique which was successfully used in the Riemannian case \cite{Gurses}.

Quite interestingly, in the literature there are similar solutions reported \cite{Chen1,Chen2} with the black hole metric configurations of the same type, however, with different torsion configurations. They are not described by the ansatz (\ref{cofB}) and (\ref{gamB}), a possibility of constructing an appropriate generalization of this ansatz will be discussed elsewhere. More recently, spherically symmetric solutions were obtained \cite{Cem1,Cem2} for the special case of the parity-even class of quadratic Poincar\'e gravity models. These configurations do not satisfy the double-duality ansatz, with the dynamical axial trace torsion field playing the role of a Maxwell--Coulomb field which gives rise to an effective Reissner--Nordstr\"om-type line element. Such solutions explicitly demonstrate that the generalized Birkhoff's theorem is not valid for the whole class of quadratic Poincar\'e gauge gravity models, see the relevant discussion in~\cite{Tartu,OPZ}. 

The final remark is as follows. There is common belief in the validity of the statement ``black holes do not have hair'' which means that there are no non-metric field configurations that satisfy the vacuum field equations and are regular across a black hole horizon. This seems to be generally true for all matter fields with an exception of the electromagnetic field. Our solutions provide a kind of counter-example to the no-hair conjecture in the sense that the corresponding dynamical torsion field is regular across the black hole horizon. The crucial role is played by the properties (\ref{invB}) which show a global regularity of the Riemann--Cartan geometry for our solutions. In other words, we have demonstrated that a black hole may have a nontrivial ``geometrical hair'' (in the form of the spacetime torsion) which does not affect the structure of a usual Kerr--de Sitter black hole. 

%%%%%%%%%%%%%%%%%%%%%%%%%%%%%%%%%%%%%%%%%%
\subsubsection*{Acknowledgments}Early discussions with Jens Boos are gratefully acknowledged. I thank the Organizers of the workshop ``Teleparallel Universes in Salamanca'' (26--28 November 2018) for the invitation and support of my~visit. This work was partially supported by the Russian Foundation for Basic Research (Grant No. 18-02-40056-mega).

\appendix
\section{Irreducible Decompositions}\label{appA}
% \unskip
%%%%%%%%%%%%%%%%%%%%%%%%%%%%%%%%%%%%%%%%%%
\subsection{Torsion}\label{appA1}
%%%%%%%%%%%%%%%%%%%%%%%%%%%%%%%%%%%%%%%%%%

The torsion 2-form can be decomposed into the three 
irreducible pieces, $T^{\alpha}={}^{(1)}T^{\alpha} + {}^{(2)}T^{\alpha} + 
{}^{(3)}T^{\alpha}$, where the torsion trace, the axial torsion, and the purely tensor torsion are defined by
\begin{eqnarray}
{}^{(2)}T^{\alpha}&=& {\frac 13}\vartheta^{\alpha}\wedge (e_\nu\rfloor 
T^\nu),\label{iT2}\\
{}^{(3)}T^{\alpha}&=& {\frac 13}e^\alpha\rfloor(T^{\nu}\wedge
\vartheta_{\nu}),\label{iT3}\\
{}^{(1)}T^{\alpha}&=& T^{\alpha}-{}^{(2)}T^{\alpha} - {}^{(3)}T^{\alpha}.
\label{iT1}
\end{eqnarray}

%%%%%%%%%%%%%%%%%%%%%%%%%%%%%%%%%%%%%%%%%%
\subsection{Curvature}\label{appA2}
%%%%%%%%%%%%%%%%%%%%%%%%%%%%%%%%%%%%%%%%%%

The Riemann--Cartan curvature 2-form is decomposed $R^{\alpha\beta} = 
\sum_{I=1}^6\,{}^{(I)}\!R^{\alpha\beta}$ into the 6 irreducible~parts 
\begin{eqnarray}
{}^{(2)}\!R^{\alpha\beta} &=& -\,{}^*(\vartheta^{[\alpha}\wedge
\overline{\Psi}{}^{\beta]}),\label{curv2}\\
{}^{(3)}\!R^{\alpha\beta} &=& -\,{\frac 1{12}}\,{}^*(\overline{X}
\,\vartheta^\alpha\wedge\vartheta^\beta),\label{curv3}\\
{}^{(4)}\!R^{\alpha\beta} &=& -\,\vartheta^{[\alpha}\wedge\Psi^{\beta]},\label{curv4}\\
{}^{(5)}\!R^{\alpha\beta} &=& -\,{\frac 12}\vartheta^{[\alpha}\wedge e^{\beta]}
\rfloor(\vartheta^\gamma\wedge X_\gamma),\label{curv5}\\
{}^{(6)}\!R^{\alpha\beta} &=& -\,{\frac 1{12}}\,X\,\vartheta^\alpha\wedge
\vartheta^\beta,\label{curv6}\\
{}^{(1)}\!R^{\alpha\beta} &=& R^{\alpha\beta} -  
\sum\limits_{I=2}^6\,{}^{(I)}R^{\alpha\beta},\label{curv1}
\end{eqnarray}
where 
\begin{eqnarray}
&X^\alpha := e_\beta\rfloor R^{\alpha\beta},\qquad X := e_\alpha\rfloor X^\alpha,\qquad
\overline{X}{}^\alpha := {}^*(R^{\beta\alpha}\wedge\vartheta_\beta),\qquad
\overline{X} := e_\alpha\rfloor \overline{X}{}^\alpha,&\label{WX2}\\
&\Psi_\alpha := X_\alpha - {\frac 14}\,\vartheta_\alpha\,X - {\frac 12}
\,e_\alpha\rfloor (\vartheta^\beta\wedge X_\beta),\qquad 
\overline{\Psi}{}_\alpha := \overline{X}{}_\alpha - {\frac 14}\,\vartheta_\alpha
\,\overline{X} - {\frac 12}\,e_\alpha\rfloor (\vartheta^\beta\wedge 
\overline{X}{}_\beta).&\label{Phia}
\end{eqnarray}

%%%%%%%%%%%%%%%%%%%%%%%%%%%%%%%%%%%%%%%%%%
\subsection{Elementary Properties}\label{appA3}
%%%%%%%%%%%%%%%%%%%%%%%%%%%%%%%%%%%%%%%%%%

Directly from the definitions (\ref{iT2}--\ref{iT1}) and (\ref{curv2}--\ref{curv1}),
one can prove the relations
\begin{eqnarray}
&T^\alpha\wedge{}^{(1)}\!T_\alpha = {}^{(1)}\!T^\alpha\wedge{}^{(1)}\!T_\alpha,&\label{T11}\\
\label{T23}
&T^\alpha\wedge{}^{(2)}\!T_\alpha = T^\alpha\wedge{}^{(3)}\!T_\alpha = {}^{(2)}\!T^\alpha\wedge{}^{(3)}\!T_\alpha,&\\
&R^{\alpha\beta}\wedge{}^{(1)}\!R_{\alpha\beta} = {}^{(1)}\!R^{\alpha\beta}\wedge{}^{(1)}
\!R_{\alpha\beta},& \label{R11}\\
&R^{\alpha\beta}\wedge{}^{(5)}\!R_{\alpha\beta} = {}^{(5)}\!R^{\alpha\beta}
\wedge{}^{(5)}\!R_{\alpha\beta},&\label{R55}\\
&R^{\alpha\beta}\wedge{}^{(2)}\!R_{\alpha\beta} = R^{\alpha\beta}\wedge{}^{(4)}\!R_{\alpha\beta} 
= {}^{(2)}\!R^{\alpha\beta}\wedge{}^{(4)}\!R_{\alpha\beta},&\label{R24} \\ 
&R^{\alpha\beta}\wedge{}^{(3)}\!R_{\alpha\beta} = R^{\alpha\beta}\wedge{}^{(6)}\!R_{\alpha\beta} 
= {}^{(3)}\!R^{\alpha\beta}\wedge{}^{(6)}R_{\alpha\beta}.&\label{R36}
\end{eqnarray}

%%%%%%%%%%%%%%%%%%%%%%%%%%%%%%%%%%%%%%%%%%
\section{de Sitter Geometry}\label{appB}
%%%%%%%%%%%%%%%%%%%%%%%%%%%%%%%%%%%%%%%%%%

The Riemannian connection is uniquely determined by the torsion-free condition (\ref{DSnoT}): 
\begin{equation}\label{gamab}
\widehat{\Gamma}{}_{\alpha\beta} = {\frac 12}\left(\widehat{e}{}_\alpha\rfloor d\widehat{\vartheta}{}_\beta - \widehat{e}{}_\beta\rfloor d\widehat{\vartheta}{}_\alpha - \widehat{\vartheta}{}^\gamma\,\widehat{e}{}_\alpha\rfloor \widehat{e}{}_\beta\rfloor d\widehat{\vartheta}{}_\gamma\right).
\end{equation}

Explicitly, we find for the components of the local Lorentz connection:
\begin{align}\label{Gds1}
\widehat{\Gamma}{}_{\hat{0}\hat{1}} &= -\,(\beta_1 + \beta_2)\widehat{\vartheta}{\,}^{\hat{0}} +
\beta_2\widehat{\vartheta}{\,}^{\hat{1}} + \alpha_1\widehat{\vartheta}{\,}^{\hat{3}}, &
\widehat{\Gamma}{}_{\hat{2}\hat{3}} &= -\,\alpha_2\widehat{\vartheta}{\,}^{\hat{0}} + 
\alpha_4\widehat{\vartheta}{\,}^{\hat{1}} - \beta_5\widehat{\vartheta}{\,}^{\hat{3}},\\
\widehat{\Gamma}{}_{\hat{0}\hat{2}} &= \alpha_3\widehat{\vartheta}{\,}^{\hat{0}} +
\beta_3\widehat{\vartheta}{\,}^{\hat{2}} + \alpha_2\widehat{\vartheta}{\,}^{\hat{3}}, &
\widehat{\Gamma}{}_{\hat{3}\hat{1}} &= \alpha_1\widehat{\vartheta}{\,}^{\hat{0}} - 
\alpha_4\widehat{\vartheta}{\,}^{\hat{2}} + \beta_4\widehat{\vartheta}{\,}^{\hat{3}},\label{Gds2}\\
\widehat{\Gamma}{}_{\hat{0}\hat{3}} &= \alpha_1\widehat{\vartheta}{\,}^{\hat{1}} - 
\alpha_2\widehat{\vartheta}{\,}^{\hat{2}} + \beta_3\widehat{\vartheta}{\,}^{\hat{3}}, &
\widehat{\Gamma}{}_{\hat{1}\hat{2}} &= -\,\alpha_3\widehat{\vartheta}{\,}^{\hat{1}} - 
\beta_4\widehat{\vartheta}{\,}^{\hat{2}} - \alpha_4\widehat{\vartheta}{\,}^{\hat{3}},\label{Gds3}
\end{align}
where we introduced the abbreviations
\begin{eqnarray}
&\alpha_1 = \sqrt{\frac {f}{\Sigma}}\,{\frac {j_0r\sin\theta}{\Sigma}},\qquad 
\alpha_2 = {\frac {1}{\sqrt{\Delta\Sigma}}}\,{\frac {j_0\cos\theta(\Delta + mr)}{\Sigma}},\qquad
\alpha_3 = \sqrt{\frac {f}{\Sigma}}\,{\frac {j_0^2r\sin\theta\cos\theta}{\Sigma}},&\label{dsa3}\\
&\alpha_4 = {\frac {1}{\sqrt{\Delta\Sigma}}}\,{\frac {j_0mr\cos\theta}{\Sigma}},\quad
\beta_1 = {\frac 12}\sqrt{\frac {\Delta}{\Sigma}}\left({\frac {\Delta'}{\Delta}}
- {\frac {\Sigma'}{\Sigma}}\right),\quad
\beta_2 = {\frac m{2\sqrt{\Delta\Sigma}}}\left(2 - r{\frac {\Delta'}{\Delta}}
- r{\frac {\Sigma'}{\Sigma}}\right),&\label{dsb2}\\
&\beta_3 = {\frac {1}{\sqrt{\Delta\Sigma}}}\,{\frac {mr^2}{\Sigma}},\quad 
\beta_4 = {\frac 1{\sqrt{\Delta\Sigma}}}\,{\frac {r(\Delta + mr)}{\Sigma}},\quad
\beta_5 = \sqrt{\frac {f}{\Sigma}}\left[\cot\theta
+ j_0^2\sin\theta\cos\theta\left({\frac {1}{\Sigma}}
- {\frac {\lambda}{f}}\right)\right].&\label{dsb5}
\end{eqnarray}

Here, the prime denotes the derivative with respect to the radial coordinate $' = \partial_r$. We split the coefficients into two groups: $\alpha_1,\dots , \alpha_4$ vanish in the absence of rotation (when $j_0 = 0$), whereas $\beta_1,\dots , \beta_5$ are always nontrivial.

\end{document}